\documentclass[12pt,preprint]{aastex}
\usepackage{epsfig,amssymb,natbib}
\bibpunct{(}{)}{;}{a}{}{,}

\newcommand{\lya}{Ly$\alpha$}

\shorttitle{Protocluster at $z=4.1$}
\shortauthors{Venemans et al.}

\begin{document}

   \title{The most distant structure of galaxies known: a
    protocluster at $z=4.1$\altaffilmark{1}}

   \author{B.P. Venemans\altaffilmark{2}, J.D. Kurk\altaffilmark{2},
   G.K. Miley\altaffilmark{2}, H.J.A. R\"ottgering\altaffilmark{2},
   W. van Breugel\altaffilmark{3}, C.L. Carilli\altaffilmark{4}, C. De
   Breuck\altaffilmark{5}, H. Ford\altaffilmark{6},
   T. Heckman\altaffilmark{6}, P. McCarthy\altaffilmark{7}, \&
   L. Pentericci\altaffilmark{8}}

\altaffiltext{1}{Based on observations carried out at the European
    Southern Observatory, Paranal, Chile; program LP167.A-0409}

\altaffiltext{2}{Sterrewacht Leiden, P.O. Box 9513, 2300 RA, Leiden,
   The Netherlands; venemans@strw.leidenuniv.nl,
   kurk@strw.leidenuniv.nl, miley@strw.leidenuniv.nl,
   rottgeri@strw.leidenuniv.nl.}

\altaffiltext{3}{Lawrence Livermore National Laboratory, P.O. Box 808,
Livermore CA, 94550, USA; wil@igpp.ucllnl.org.}

\altaffiltext{4}{NRAO, P.O. Box 0, Socorro NM, 87801, USA;
ccarilli@aoc.nrao.edu.}

\altaffiltext{5}{Institut d'Astrophysique de Paris, 98bis Boulevard
Arago, 75014, Paris, France; debreuck@iap.fr.}

\altaffiltext{6}{Dept. of Physics \& Astronomy, The Johns Hopkins
University, 3400 North Charles Street, Baltimore MD, 21218-2686, USA;
ford@pha.jhu.edu, heckman@pha.jhu.edu.}

\altaffiltext{7}{The Observatories of the Carnegie Institution of
Washington, 813 Santa Barbara Street, Pasadena CA, 91101, USA;
pmc2@ociw.edu.}

\altaffiltext{8}{Max-Planck-Institut f\"ur Astronomie, K\"onigstuhl
17, D-69117, Heidelberg, Germany; laura@mpia-hd.mpg.de.}

\newpage

\begin{abstract}
    Imaging and spectroscopy with the Very Large Telescope have
    revealed 20 \lya\ emitters within a projected distance of 1.3 Mpc
    and 600 km s$^{-1}$ of the luminous radio galaxy TN J1338--1942 at
    $z=4.1$. Compared to the field density of \lya\ emitters, this
    implies an overdensity on the order of 15. The structure has a
    projected size of at least 2.7 Mpc $\times$ 1.8 Mpc and a velocity
    dispersion of 325 km s$^{-1}$, which makes it the most distant
    structure known. Using the galaxy overdensity and assuming a bias
    parameter $b$ = 3 -- 5, the mass is estimated to be $\sim 10^{15}$
    $M_{\sun}$. The radio galaxy itself is surrounded by an uniquely
    asymmetric \lya\ halo. Taken together with our previous data on
    PKS 1138--262 at $z \sim 2.16$, these results suggest that
    luminous radio sources are excellent tracers of high density
    regions in the early Universe, which evolve into present-day
    clusters. The statistics of bright radio sources and of
    concentrations in the Lyman break galaxy population are consistent
    with the picture that each of those concentrations harbours an
    active or passive luminous radio source.
\end{abstract}

    \keywords{galaxies: active --- galaxies: clusters: general ---
    galaxies: evolution --- cosmology: observations --- cosmology: early
    Universe}

\section{Introduction}

Studies of high redshift (proto)clusters of galaxies ($z > 2$) can
directly constrain theories of galaxy evolution and cosmological
models \citep[e.g.][]{bah98}, but the detection of (proto)clusters at
these redshifts using conventional optical and X-ray techniques is
difficult. High redshift radio galaxies (HzRGs, $z > 2$) can help:
they are among the most evolved and most massive galaxies in the early
Universe \citep{deb02} and are most likely located in dense
(proto)cluster environments
\citep[e.g.,][]{lef96,pas96,rot96,car97,oji97,pen00b}. In addition,
HzRGs have properties that would be expected of forming central
cluster galaxies. Their extremely clumpy morphologies revealed by {\em
Hubble Space Telescope} images \citep{pen99} are strikingly similar to
simulations of forming brightest cluster galaxies, based on
hierarchical models (e.g., Arag\'on-Salamanca, Baugh, \& Kauffmann
1998).  \nocite{ara98}

A pilot project on the Very Large Telescope (VLT) to search for
\lya-emitting galaxies around the clumpy radio galaxy PKS 1138--262,
resulted in the discovery of 14 galaxies and a QSO at approximately
the same redshift as the radio galaxy \citep{kur00,pen00a}. If the
structure found is virialized, the total mass of the protocluster
would be $10^{14}$ $M_{\sun}$. Motivated by this result, we started a
large program at the VLT to search for forming clusters
(protoclusters) around HzRGs at redshifts 2 and higher. The first
radio galaxy field we observed was TN J1338--1942 at a redshift of 4.1
\citep{deb99}. This HzRG is one of the brightest and most luminous in
\lya\ known. Both its \lya\ profile and radio structure are very
asymmetric \citep{deb99}, which indicates strong interaction with
dense gas, and the rest-frame radio luminosity is comparable to that
of the most luminous 3CR sources (P$_{178\rm MHz} \simeq 4 \times
10^{35}$ erg s$^{-1}$ Hz$^{-1}$ sr$^{-1}$).  Here we report on the
discovery of a substantial overdensity of \lya\ emitters around this
radio galaxy at $z \sim 4.1$.\footnote{Throughout this Letter,
magnitudes are in the AB system and a $\Lambda$-dominated cosmology
with H$_0 = 65$ km s$^{-1}$ Mpc$^{-1}$, $\Omega_{M} = 0.3$, and
$\Omega_{\Lambda} = 0.7$ is assumed.}

\section{Observations and candidate selection}

\subsection{VLT imaging and selection of candidate \lya\ emitters}

We carried out narrow- and broadband imaging on 2001 March 25 and 26
with the 8.2 m VLT Kueyen (UT2), using the imaging mode of the FOcal
Reducer/low dispersion Spectrograph 2 (FORS2). At $z
= 4.10$ the \lya\ emission line is redshifted to 6202 \AA, which falls
in our custom narrowband filter with a central wavelength of 6195
\AA\ and FWHM of 60 \AA. The broadband R filter had a central
wavelength of 6550 \AA\ and FWHM of 1650 \AA. The detector was a
SiTE CCD with 2048$\times$2048 pixels, with a scale of 0\farcs2 per
pixel and a field of view of 6\farcm8$\times$6\farcm8. We took 18
separate 1800 s exposures and one 900 s exposure in the narrow band
and 21 exposures of 300 s in $R$, shifted by $\sim15$\arcsec\ with
respect to each other to minimize flat-fielding problems and to handle
cosmic rays. The nights were photometric and the average seeing in
both narrowband and broadband images was 0\farcs65. The 1 $\sigma$
limiting magnitude per square arcsecond was 28.6 for the narrowband
and 29.2 for the broadband image. For the flux calibration the
spectrophotometric standard star GD 108 \citep{oke90} was used. The
final images have sizes of 6\farcm4 $\times$ 6\farcm2 (39.7
arcmin$^2$). The total volume probed at $z = 4.1$ by our narrowband
filter is 7315 Mpc$^3$.

NTT imaging data, taken on 1998 April 27 and 29 under nonphotometric
conditions with SuSI2, were also used to provide $B$- and $I$-band
magnitudes for candidate emitters where possible. The 1 $\sigma$
limiting magnitude per square arcsecond was about 27.8 in both bands.

For the detection and photometry of objects in the images, we used the
object detection and classification program SExtractor \citep{ber96}.
Detected objects had at least 9 connected pixels with values larger
than the rms sky noise on the narrowband image. A photometric analysis
was then carried out on both the narrowband and the broadband
image. In total 2407 objects were extracted.  Based on the statistics
of our detection of \lya\ emitters around PKS 1138--262, we selected
objects with a rest-frame equivalent width (EW$_0)$ greater than 15
\AA\ (or $m_{\rm BB} -$ $m_{\rm NB} > 0.84$) and significance $\Sigma
> 3$ as candidate \lya-emitting galaxies, with $\Sigma$ the ratio of
continuum-subtracted counts in the narrowband to the combined noise in
the broadband and narrowband \citep{bun95}. Of the 2407 objects, 34
objects satisfied these criteria, including the radio galaxy. Of these
34 objects, 31 were detected in the $R$-band image, 9 were detected in
the NTT $I$ image and 5 in the NTT $B$ image. The 5 objects with a
detection in B have a color of $B-R$ $\sim$ 0--1, inconsistent with
the colors expected for galaxies at $z=4$ \citep{ste99}. Three of
these objects are located in the halo of the radio galaxy, which
affects the narrowband photometry. The other two are likely to be
foreground objects with another line falling in the narrowband filter,
e.g.\ [\ion{O}{3}] $\lambda5007$ or [\ion{O}{2}] $\lambda
3727$. Excluding the radio galaxy, the resulting 28 objects were our
\lya-emitting candidates for follow-up spectroscopy.

\subsection{VLT spectroscopy}

On 2001 May 20, 21, and 22 we carried out spectroscopy using FORS2 in
the mask multiobject spectroscopy mode with standard resolution. The
nights were photometric with an average seeing of 1\arcsec. The
spectra were obtained with the 600RI grism with a dispersion of 1.32
\AA\ pixel$^{-1}$ and a wavelength range from 5300 \AA\ to 8000
\AA. This grism was preferred because of its high throughput (peak
efficiency is 87\%). Two different masks were used to observe 23 of the
candidate \lya\ emitters and the radio galaxy, with slit sizes of
1\arcsec, resulting in a resolution of 6 \AA, which corresponds to
290 km s$^{-1}$ at $z = 4.1$. The total exposure time was 31,500
s for the first mask and 35,100 s for the second mask. Wavelength
calibration was obtained from exposures of He, Ar, Ne, and HgCd
arc-lamps. The accuracy of the wavelength calibration was better than
0.05 \AA. For the flux calibration, long slit exposures with a
5\arcsec\ slit were used of the spectrophotometric standard stars
EG 274 and LTT 7987 \citep{sto83,bal84}.

\section{Results}

\subsection{Line emitters in the field}

\subsubsection{Line identifications}

\begin{figure}
\plotone{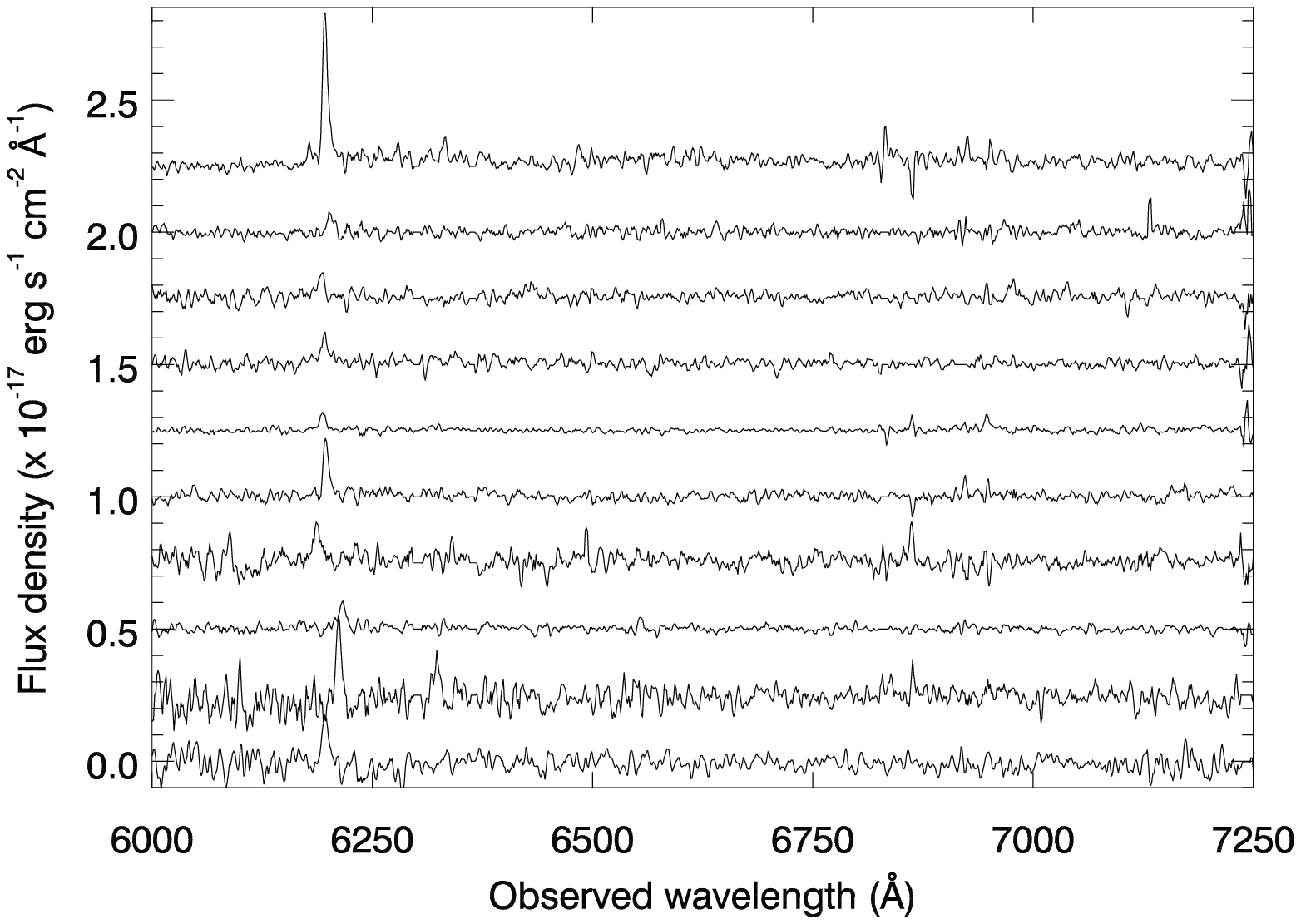}
\caption{Spectra of 10 of the 20 confirmed \lya\ emitters. For
clarity, each spectrum is offset by multiples of $2.5 \times 10^{-18}$
erg s$^{-1}$ cm$^{-2}$ \AA$^{-1}$. \label{10specs}}
\end{figure}

Of the 23 candidates observed, 20 show an emission line with a peak
between $\lambda = 6187$ \AA\ and $\lambda = 6216$ \AA.  The
signal-to-noise of these lines is at least 10. Ten randomly selected
spectra from our sample are shown in Figure \ref{10specs}. Two
nondetections were very faint ($m_{\rm NB} = 25.9$ and $m_{\rm NB} = 26.1$
respectively). The third nondetection had a very low surface
brightness. The success rate of our selection criteria was therefore
87\%.

The first question is whether the detected lines are indeed due to
\lya\ at the redshift of the radio galaxy.  Intervening neutral
hydrogen will absorb emission blueward of the Lya line
\citep{ste99}. This discontinuity of the continuum over the Lya
emission line is observed in one of the spectra, but the continuum
emission of the other emitters was too faint to be detected (R $\sim
27$). Identification of the lines with [\ion{O}{3}] $\lambda5007$ at
$z \sim 0.24$ can be excluded, because of the lack of confirming lines
[\ion{O}{3}] $\lambda 4959$ and H$\beta$ in the spectra. The position
of the emitter within the slit mask determines the wavelength coverage
of the resulting spectrum. In 9 cases this coverage was suitable to
exclude the identification with [\ion{O}{2}] $\lambda 3727$ at $z \sim
0.66$ on similar grounds. The detected emitters are anyway unlikely to
be foreground [\ion{O}{2}] galaxies. First of all, if one of the lines
would be [\ion{O}{2}], then the rest-frame equivalent width would be
at least 70 \AA. A survey of nearby field galaxies, conducted by
\citet{jan00}, gives a mean EW$_0$ of the [\ion{O}{2}] line of $\sim
30$ \AA\ for galaxies with $M_B = -16$, which roughly compares to R
$\approx 27$ at $z = 0.66$. Only 2 galaxies out of 159 galaxies with
[\ion{O}{2}] in emission have EW$_0$([\ion{O}{2}]) $> 70$ \AA. The
number of [\ion{O}{2}] emitters expected in our field, using another
study \citep{hog98}, is $\sim 7$. Again, only a few percent of the
[\ion{O}{2}] emitters observed by Hogg et al.\ have an EW$_0 > 70$
\AA. Therefore, from our sample of 20 emitters, $< 1$ is expected to
be an [\ion{O}{2}] emitter.

\begin{figure}
\plotone{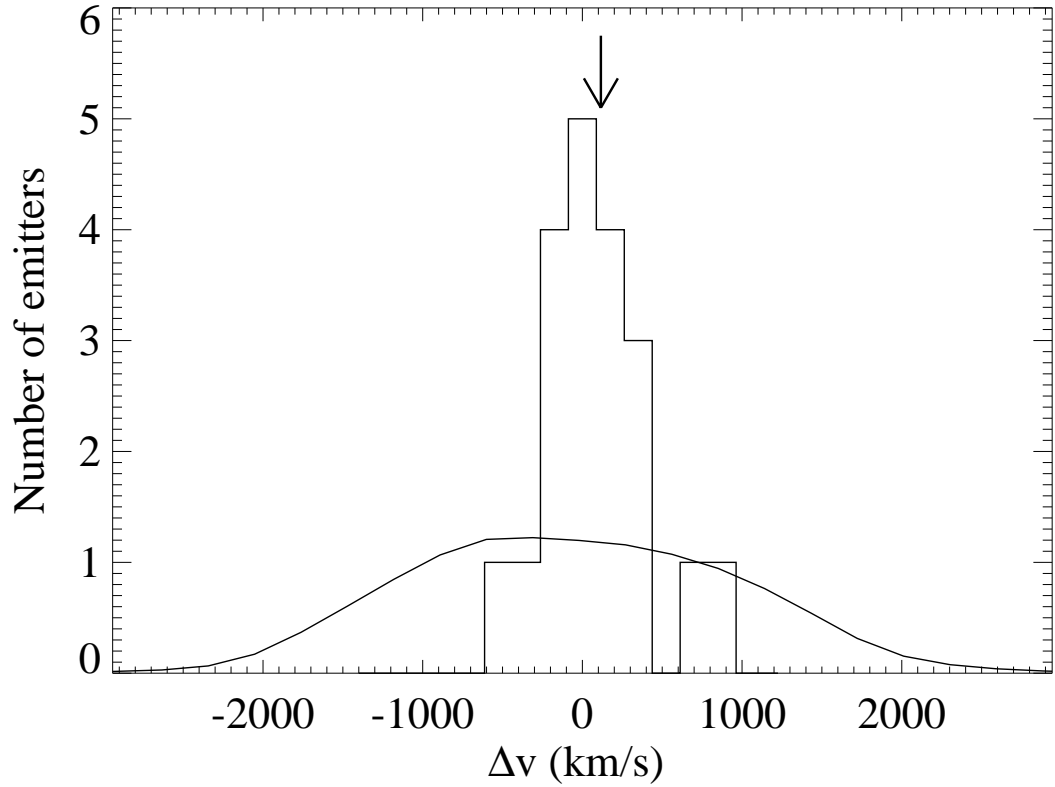}
\caption{Velocity distribution of the \lya\ emitters. The bin size is
175 km s$^{-1}$ and the median of the velocities is taken as zero
point. The velocity of the radio galaxy, corrected for absorption
\citep{deb99}, is indicated by an arrow. The normalized transmission
curve of the narrowband filter is also plotted. Note that the
velocity distribution of the detected emitters is substantially
narrower than the filter width and centered within 200 km s$^{-1}$ of
the redshift of the radio galaxy. \label{vdist}}
\end{figure}

Additional evidence that the observed lines are predominantly \lya\
lines at $z \sim 4.1$ associated with the radio galaxy is provided by
their velocity distribution, which has a dispersion of $326 \pm 73$ km
s$^{-1}$ and a FWHM of $768 \pm 172$ km s$^{-1}$ (Fig.\ \ref{vdist}),
which is 4 times smaller than the FWHM of the narrowband
filter. Further, the velocity distribution peaks within 200 km
s$^{-1}$ of the radio galaxy \lya\ peak, corrected for absorption
\citep{deb99}.

For all these reasons we interpret the observed emission lines as
\lya. To estimate the redshifts, flux densities, and widths (FWHM) of
the lines, a Gaussian function was fitted to each of the 
one-dimensional spectra. Details will be provided in a future paper
(B.P.\ Venemans et al., in preparation).

\subsubsection{Significance and properties of the overdensity}

The next question to be addressed is to what extent the statistics of
our detections represent a significant overdensity of galaxies in the
field. A ``blank-field'' study of \lya\ emitters at approximately the
same redshift is the Large-Area Lyman Alpha survey \citep[LALA survey,
][]{rho00}. Preliminary results of this survey indicate a number
density of $4000 \pm 460$ deg$^{-2}$ $\Delta z^{-1}$ for objects with
EW$_0 \gtrsim 15$\AA\ and line + continuum $> 2.6 \times 10^{-17}$ erg
s$^{-1}$ cm$^{-2}$. For our field the second
criterion corresponds to $m_{\rm NB} < 24.55$ and we expect $2.3 \pm 0.3$ such
\lya\ emitters within our probed volume. However, 9 of the confirmed
emitters satisfy the above criteria (including the radio galaxy). In
our cosmology the comoving volume density of the LALA survey is
$n_{LALA} = (3.1 \pm 0.4) \times 10^{-4}$ Mpc$^{-3}$. The volume
density in our field is $n_{1338} = 9 / 7315$ Mpc$^{-3} = (12 \pm 4) \times
10^{-4}$ Mpc$^{-3}$. The difference in number density is
$n_{1338}$/$n_{LALA} = 4.0 \pm 1.4$. However, the FWHM of the
velocity distribution is approximately 4 times smaller than the FWHM
of the filter (see Fig.\ \ref{vdist}). This implies that our radio
galaxy field is overdense in \lya\ emitters by a factor of 15 compared
with a blank field.

\begin{figure}
\plotone{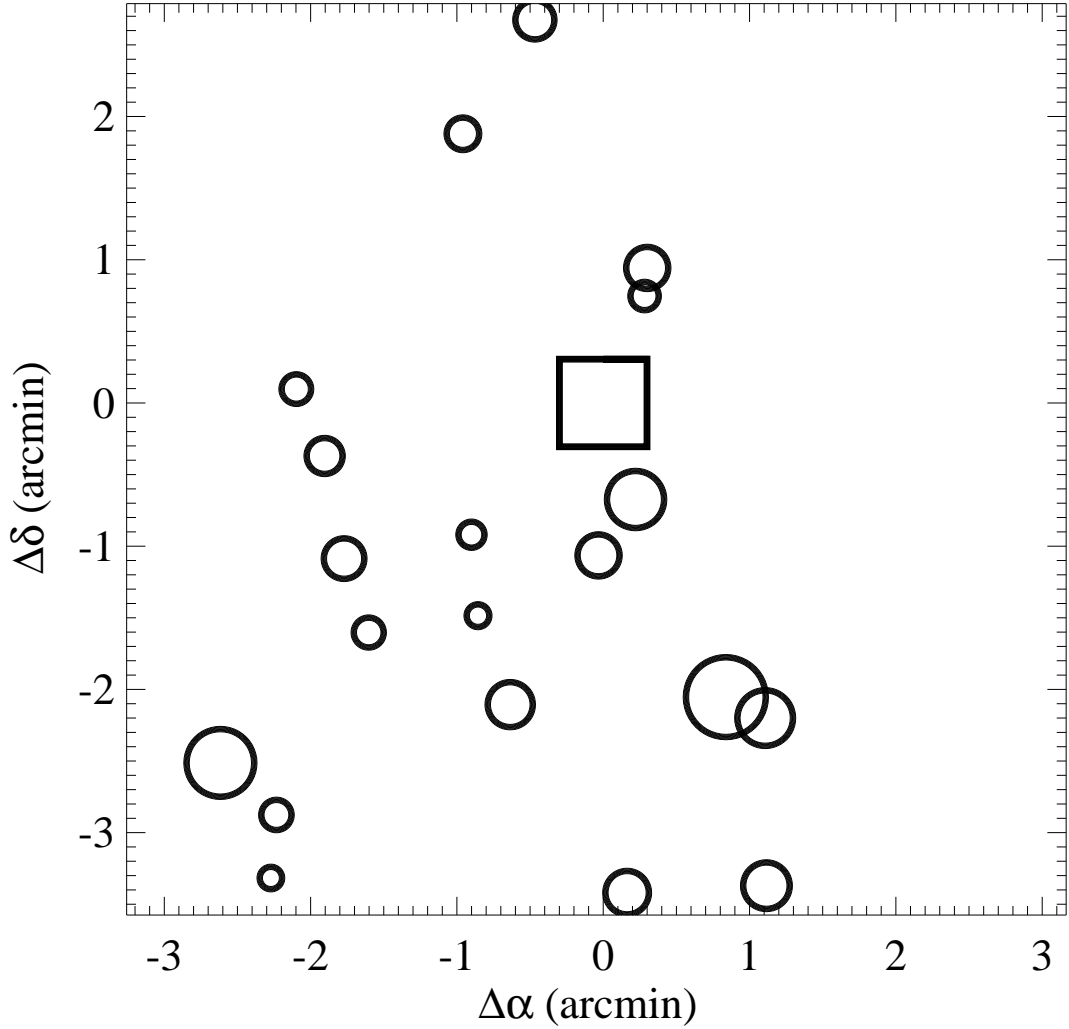}
\caption{Spatial distribution of the 20 confirmed \lya\ emitters at $z
\sim 4.1$ ({\em circles}) and the radio galaxy ({\em square}). The size of the
circles is scaled according to the \lya\ flux of the object, ranging
between 0.3 and 4.1 $\times 10^{-17}$ erg s$^{-1}$ cm$^{-2}$. The
structure appears to be bound in the northwest of the image and
unbound in the south. Note that the radio galaxy is not centered in
the galaxy distribution. \label{sdist}}
\end{figure}

The spatial distribution of the emitters is not homogeneous (Fig.\
\ref{sdist}). The structure appears to be bound in the northwest but
our FOV is not large enough to show a boundary in the south. The size
of the structure is therefore at least 6\arcmin $\times$ 4\arcmin,
which corresponds to greater than 2.7 Mpc $\times$ 1.8 Mpc. Remarkably, the
radio galaxy, probably the most massive system in the structure, does
not appear to be at the center of the protocluster, in disagreement
with models for the formation of dominant cluster galaxies
\citep[e.g.,][]{wes94}.

\subsection{Radio galaxy halo}

\begin{figure}
\plotone{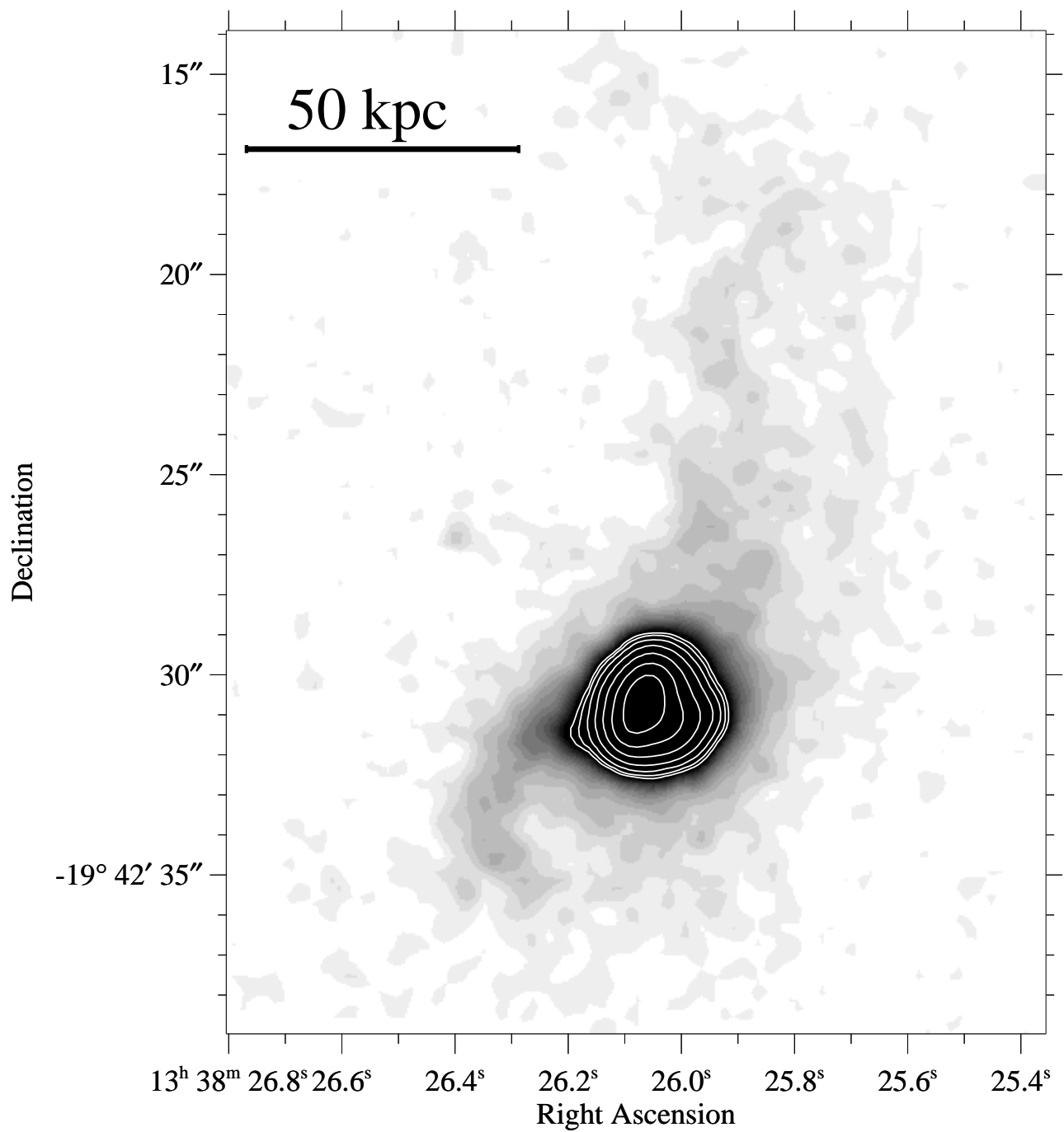}
\caption{Continuum-subtracted \lya\ image of the radio galaxy
  halo. The contours represent the \lya\ flux density in the center of
  the halo. The surface brightness ranges from $\sim 0.07$ to
  $1 \times 10^{-17}$ erg s$^{-1}$ cm$^{-2}$ arcsec$^{-2}$. The
  low-brightness halo is extended to the northwest, pointing away from
  the overdensity structure. This \lya\ halo is the most asymmetric
  radio galaxy halo known. \label{halo}}
\end{figure}

The radio galaxy is located close to the apparent north-west boundary
of the galaxy overdensity structure. The radio emission is dominated
by two components, separated by 5\farcs5 \citep{deb99}. The brightest
component coincides with the optical emission, while the other is in
the south-east, towards the center of the galaxy overdensity.
A spectacular feature of the radio galaxy, visible in the
narrowband image, is the large \lya\ halo (Fig.\ \ref{halo}). Although
giant \lya\ halos are a common feature of HzRGs, they are usually
fairly symmetrically extended around the radio galaxy. In the case of
TN J1338--1942, the halo is highly asymmetric and extends for
$\sim15$\arcsec\ (110 kpc) to the north-west, i.e., away from the
center of the overdensity structure. In its narrowness and asymmetry
the TN J1338--1942 halo is unique among known distant radio
galaxies. Possible mechanisms to produce this structure include
cooling flows in colliding sub-structures and buoyancy effects
\citep{gis76} and will be considered in a subsequent paper.

\section{Discussion}

\subsection{Nature of the overdensity}

Could the structure that we have detected be a protocluster at $z
\sim 4.1$ that will evolve into a rich cluster of galaxies in the
local Universe? At $z=4.1$ the Universe is $\sim 1.6$ Gyr old, too
short for the structure to have virialized since the mean crossing
time for galaxies at the observed velocity dispersion is at least 4
Gyr. Thus the observed velocities are probably infall velocities of
the galaxies accreting onto a large overdensity. The total mass of
this structure can be estimated by using $M = \overline{\rho} V
(1+\delta_m)$ with $\overline{\rho}$ the mean density of the Universe
and $\delta_m$ the mass overdensity within our volume $V$
\citep{ste98}. The mass overdensity is related to the galaxy
overdensity $\delta_{\rm gal}$ through $1 + b \delta_m = C (1 +
\delta_{\rm gal})$, where $C$ takes into account the redshift space
distortions caused by peculiar velocities and $b$ is the bias
parameter. From the statistics of redshift ``spikes'', \citet{ste98}
argue that $b \gtrsim 4$. Taking $b$ in the range 3--5 and assuming
that the structure is just breaking away from the Hubble expansion,
$\delta_m$ is estimated to be 0.5--2.3, giving a mass of our structure
of (1--2) $\times 10^{15}$ $M_{\sun}$. This is comparable to the mass of
the Lyman break galaxy overdensity (spike) found by Steidel et
al.\ and to that of the Coma cluster \citep[e.g.,][]{fus94}.

\subsection{Relation to overdensity spikes}

It is instructive to compare the number of luminous radio sources at
$z \sim 3$ to the number of redshift spikes. \citet{ste98}
estimate that about one velocity spike is detected in each 9\arcmin
$\times$ 18\arcmin\ field in their spectroscopic survey, corresponding
to $9 \times 10^5$ redshift spikes in the whole sky. Their survey is
sensitive to redshifts between 2.7 and 3.4, corresponding to a cosmic
evolution time of 0.6 Gyr.

How many luminous radio sources are there at $z \sim 3$? Using the
pure luminosity evolution model of \citet{dun90} to describe the steep
spectrum radio luminosity function, we estimate that in the redshift
range $2.7 < z < 3.4$ there should be $\sim 1.2 \times 10^4$ radio
sources with luminosities exceeding $10^{33}$ erg s$^{-1}$ Hz$^{-1}$
sr$^{-1}$ at 2.7 GHz (``Cygnus A type'' radio sources:
P$_{2.7\rm GHz}$(Cygnus A) $\simeq 2 \times 10^{33}$ erg s$^{-1}$
Hz$^{-1}$ sr$^{-1}$, Becker, White, \& Edwards 1991). \nocite{bec91}
Assuming that HzRGs are only once active for $10^7$ yr \citep{blu99},
we expect the number of (previously) active radio sources in this
redshift range to be $\sim 7 \times 10^5$.

Hence, the luminosity functions and lifetimes of luminous radio
sources are consistent with every velocity ``spike'' in the space
densities of Lyman break galaxies being associated with a massive
galaxy that has been or will become a luminous radio source once. 
Note that \citet{wes94} presented similar statistical evidence to
argue that distant powerful radio galaxies are the precursors of cD
galaxies at the centers of galaxy clusters.

\section{Conclusion}

We have found a structure of 20 \lya-emitting galaxies around the
high redshift radio galaxy TN J1338--1942. The overdensity of this
protocluster is on the order of 15 compared to field samples. Our
results demonstrate that by $z = 4.1$, megaparsec-scale structure had
already formed.

Together with our previous data, this implies that the most luminous
radio sources are tracers of regions of galaxy overdensity in the
early Universe. The estimated masses of $10^{14}$ -- $10^{15}$
$M_{\sun}$ are consistent with the overdensities being ancestors
of rich clusters of galaxies in the local Universe.

\begin{acknowledgements}
We thank the staff on Paranal, Chile and Gero Rupprecht at ESO for
their splendid support, and William Grenier of Andover Corporation for
his efforts in ensuring the narrowband filter was manufactured in
time for the observations. We acknowledge useful discussions with
P.\ Martin (University of Toronto) and M.\ Davis (University of
California at Berkeley) on radio galaxy halos and
protoclusters, and their possible role in constraining cosmological
models. We thank the referee, Adam Stanford, for improving this Letter. The
work by W.\ v.\ B.\ was performed under the auspices of the US Department of
Energy, National Nuclear Security Administration by the University of
California, Lawrence Livermore National Laboratory under contract
No. W-7405-Eng-48. The NRAO is operated by associated universities
inc., under cooperative agreement with the NSF. This work was supported
by the European Community Research and Training Network ``The Physics
of the Intergalactic Medium''.
\end{acknowledgements}

\newpage

\newpage

\begin{thebibliography}{31}
\expandafter\ifx\csname natexlab\endcsname\relax\def\natexlab#1{#1}\fi

\bibitem[{{Aragon-Salamanca} {et~al.}(1998){Aragon-Salamanca}, {Baugh}, \&
  {Kauffmann}}]{ara98}
{Arag\'on-Salamanca}, A., {Baugh}, C.~M., \& {Kauffmann}, G. 1998, \mnras, 297,
  427

\bibitem[{{Bahcall} \& {Fan}(1998)}]{bah98}
{Bahcall}, N.~A. \& {Fan}, X. 1998, \apj, 504, 1

\bibitem[{{Baldwin} \& {Stone}(1984)}]{bal84}
{Baldwin}, J.~A. \& {Stone}, R.~P.~S. 1984, \mnras, 206, 241

\bibitem[{{Becker} {et~al.}(1991){Becker}, {White}, \& {Edwards}}]{bec91}
{Becker}, R.~H., {White}, R.~L., \& {Edwards}, A.~L. 1991, \apjs, 75, 1

\bibitem[{{Bertin} \& {Arnouts}(1996)}]{ber96}
{Bertin}, E. \& {Arnouts}, S. 1996, \aaps, 117, 393

\bibitem[{{Blundell} \& {Rawlings}(1999)}]{blu99}
{Blundell}, K.~M. \& {Rawlings}, S. 1999, \nat, 399, 330

\bibitem[{{Bunker} {et~al.}(1995){Bunker}, {Warren}, {Hewett}, \&
  {Clements}}]{bun95}
{Bunker}, A.~J., {Warren}, S.~J., {Hewett}, P.~C., \& {Clements}, D.~L. 1995,
  \mnras, 273, 513

\bibitem[{{Carilli} {et~al.}(1997){Carilli}, {R\"{o}ttgering}, {van Ojik},
  {Miley}, \& {van Breugel}}]{car97}
{Carilli}, C.~L., {R\"{o}ttgering}, H.~J.~A., {van Ojik}, R., {Miley}, G.~K.,
  \& {van Breugel}, W.~J.~M. 1997, \apjs, 109, 1

\bibitem[{{De Breuck} {et~al.}(1999){De Breuck}, {van Breugel}, {Minniti},
  {Miley}, {R{\" o}ttgering}, {Stanford}, \& {Carilli}}]{deb99}
{De Breuck}, C., {van Breugel}, W., {Minniti}, D., {Miley}, G.~K., {R{\"
  o}ttgering}, H.~J.~A., {Stanford}, S.~A., \& {Carilli}, C. 1999, \aap, 352, L51

\bibitem[{{De Breuck} {et~al.}(2002){De Breuck}, {van Breugel}, {Stanford},
  {R{\" o}ttgering}, \& {Miley}}]{deb02}
{De Breuck}, C., {van Breugel}, W., {Stanford}, S.~A., {R{\" o}ttgering}, H.~J.~A., {Miley}, G.~M., {Stern}, D.\ 2002, \aj, in press (astro-ph/0109540)

\bibitem[{{Dunlop} \& {Peacock}(1990)}]{dun90}
{Dunlop}, J.~S. \& {Peacock}, J.~A. 1990, \mnras, 247, 19

\bibitem[{{Fusco-Femiano} \& {Hughes}(1994)}]{fus94}
{Fusco-Femiano}, R. \& {Hughes}, J.~P. 1994, \apj, 429, 545

\bibitem[{{Gisler}(1976)}]{gis76}
{Gisler}, G.~R. 1976, \aap, 51, 137

\bibitem[{{Hogg} {et~al.}(1998){Hogg}, {Cohen}, {Blandford}, \&
  {Pahre}}]{hog98}
{Hogg}, D.~W., {Cohen}, J.~G., {Blandford}, R., \& {Pahre}, M.~A. 1998, \apj,
  504, 622

\bibitem[{{Jansen} {et~al.}(2000){Jansen}, {Fabricant}, {Franx}, \&
  {Caldwell}}]{jan00}
{Jansen}, R.~A., {Fabricant}, D., {Franx}, M., \& {Caldwell}, N. 2000, \apjs,
  126, 331

\bibitem[{{Kurk} {et~al.}(2000){Kurk}, {R{\" o}ttgering}, {Pentericci},
  {Miley}, {van Breugel}, {Carilli}, {Ford}, {Heckman}, {McCarthy}, \&
  {Moorwood}}]{kur00}
{Kurk}, J.~D., et al.\ 2000, \aap, 358, L1

\bibitem[{{Le F\`evre} {et~al.}(1996){Le F\`evre}, {Deltorn}, {Crampton}, \&
  {Dickinson}}]{lef96}
{Le F\`evre}, O., {Deltorn}, J.~M., {Crampton}, D., \& {Dickinson}, M. 1996,
  \apjl, 471, L11

\bibitem[{{Oke}(1990)}]{oke90}
{Oke}, J.~B. 1990, \aj, 99, 1621

\bibitem[{{Pascarelle} {et~al.}(1996){Pascarelle}, {Windhorst}, {Driver},
  {Ostrander}, \& {Keel}}]{pas96}
{Pascarelle}, S.~M., {Windhorst}, R.~A., {Driver}, S.~P., {Ostrander}, E.~J.,
  \& {Keel}, W.~C. 1996, \apjl, 456, L21

\bibitem[{{Pentericci} {et~al.}(2000{\natexlab{a}}){Pentericci}, {Kurk}, {R{\"
  o}ttgering}, {Miley}, {van Breugel}, {Carilli}, {Ford}, {Heckman},
  {McCarthy}, \& {Moorwood}}]{pen00a}
{Pentericci}, L., et al.\ 2000{\natexlab{a}}, \aap, 361, L25

\bibitem[{{Pentericci} {et~al.}(1999){Pentericci}, {R{\" o}ttgering}, {Miley},
  {McCarthy}, {Spinrad}, {van Breugel}, \& {Macchetto}}]{pen99}
{Pentericci}, L., {R{\" o}ttgering}, H.~J.~A., {Miley}, G.~K., {McCarthy}, P.,
  {Spinrad}, H., {van Breugel}, W.~J.~M., \& {Macchetto}, F. 1999, \aap, 341,
  329

\bibitem[{{Pentericci} {et~al.}(2000{\natexlab{b}}){Pentericci}, {Van Reeven},
  {Carilli}, {R{\" o}ttgering}, \& {Miley}}]{pen00b}
{Pentericci}, L., {Van Reeven}, W., {Carilli}, C.~L., {R{\" o}ttgering},
  H.~J.~A., \& {Miley}, G.~K. 2000{\natexlab{b}}, \aaps, 145, 121

\bibitem[{{Rhoads} {et~al.}(2000){Rhoads}, {Malhotra}, {Dey}, {Stern},
  {Spinrad}, \& {Jannuzi}}]{rho00}
{Rhoads}, J.~E., {Malhotra}, S., {Dey}, A., {Stern}, D., {Spinrad}, H., \&
  {Jannuzi}, B.~T. 2000, \apjl, 545, L85

\bibitem[{{R\"{o}ttgering} {et~al.}(1996){R\"{o}ttgering}, {West}, {Miley}, \&
  {Chambers}}]{rot96}
{R\"{o}ttgering}, H.~J.~A., {West}, M.~J., {Miley}, G.~K., \& {Chambers}, K.~C.
  1996, \aap, 307, 376

\bibitem[{{Steidel} {et~al.}(1998){Steidel}, {Adelberger}, {Dickinson},
  {Giavalisco}, {Pettini}, \& {Kellogg}}]{ste98}
{Steidel}, C.~C., {Adelberger}, K.~L., {Dickinson}, M., {Giavalisco}, M.,
  {Pettini}, M., \& {Kellogg}, M. 1998, \apj, 492, 428

\bibitem[{{Steidel} {et~al.}(1999){Steidel}, {Adelberger}, {Giavalisco},
  {Dickinson}, \& {Pettini}}]{ste99}
{Steidel}, C.~C., {Adelberger}, K.~L., {Giavalisco}, M., {Dickinson}, M., \&
  {Pettini}, M. 1999, \apj, 519, 1

\bibitem[{{Stone} \& {Baldwin}(1983)}]{sto83}
{Stone}, R.~P.~S. \& {Baldwin}, J.~A. 1983, \mnras, 204, 347

\bibitem[{{van Ojik} {et~al.}(1997){van Ojik}, {R\"{o}ttgering}, {Miley}, \&
  {Hunstead}}]{oji97}
{van Ojik}, R., {R\"{o}ttgering}, H.~J.~A., {Miley}, G.~K., \& {Hunstead},
  R.~W. 1997, \aap, 317, 358

\bibitem[{{West}(1994)}]{wes94}
{West}, M.~J. 1994, \mnras, 268, 79

\end{thebibliography}
\end{document}